\documentclass[12pt,a4paper]{article}

\usepackage[truedimen,margin=30mm]{geometry}

\usepackage{amssymb}
\usepackage{amsmath}
\usepackage{amsthm}
\usepackage[dvipdfmx]{graphicx}
\usepackage{bbm}
\usepackage{appendix}
\usepackage{amsfonts}
\usepackage{multirow}
\usepackage[flushleft]{threeparttable}
\usepackage{booktabs}
\usepackage{threeparttable}
\usepackage{dcolumn}
\usepackage{siunitx}
\usepackage{tabularx}
\usepackage{comment}
\usepackage{url}
\usepackage{tabularx}
\usepackage{bm}
\usepackage{array}
\usepackage{longtable}
\usepackage[unicode,colorlinks=true,allcolors=blue]{hyperref}
\usepackage{lmodern}
\usepackage{natbib}
\usepackage{mathrsfs}
\usepackage{ascmac}
\usepackage{setspace}
\usepackage{algorithm}
\usepackage{algpseudocode}
\usepackage{color}
\usepackage{times}
\usepackage{comment}

\usepackage{titlesec}
\titleformat*{\section}{\large\bfseries}
\titleformat*{\subsection}{\it}

\title{{\bf Bayesian Time-Varying Meta-Analysis via Hierarchical Mean-Variance Random-effects Models }}

\date{}

\begin{document}

\maketitle
\doublespacing

\vspace{-1.5cm}
\begin{center}
Kohsuke Kubota$^1$$^2$, Shonosuke Sugasawa$^3$, Keiichi Ochiai$^1$ and Takahiro Hoshino$^3$

\medskip

\medskip
\noindent
$^1$NTT DOCOMO, INC.\\
$^2$Graduate School of Economics, Keio University\\
$^3$Faculty of Economics, Keio University\\
\end{center}

\vspace{0.2cm}
\begin{center}
{\bf \large Abstract}
\end{center}

\vspace{-0cm}
Meta-analysis is widely used to integrate results from multiple experiments to obtain generalized insights. 
Since meta-analysis datasets are often heteroscedastic due to varying subgroups and temporal heterogeneity arising from experiments conducted at different time points, the typical meta-analysis approach, which assumes homoscedasticity, fails to adequately address this heteroscedasticity among experiments.
This paper proposes a new Bayesian estimation method that simultaneously shrinks estimates of the means and variances of experiments using a hierarchical Bayesian approach while accounting for time effects through a Gaussian process. 
This method connects experiments via the hierarchical framework, enabling ``borrowing strength'' between experiments to achieve high-precision estimates of each experiment's mean. 
The method can flexibly capture potential time trends in datasets by modeling time effects with the Gaussian process. 
We demonstrate the effectiveness of the proposed method through simulation studies and illustrate its practical utility using a real marketing promotions dataset.


\section{Introduction}
\label{sec:intro}
Meta-analysis is used to integrate results from multiple experiments to obtain generalized insights in fields such as epidemiology~\citep{kleinbaum1991epidemiologic}, economics~\citep{stanley2012meta}, and engineering~\citep{sohn1999meta}.
In meta-analyses, heterogeneous effects among experiments are modeled using meta-regression.
To address between-study heterogeneity from subgroups with different true effects, hierarchical multilevel models~\citep[e.g.][]{pastor2018on}, particularly hierarchical Bayesian models incorporating random effects~\citep[e.g.][]{higgins2009re, browne2017works}, are commonly used.
While these models typically allow individual study effects to vary, they generally consolidate what could be multiple, distinct between-study variances into a single summary parameter~(e.g., $\tau^2$). 
This approach inherently assumes that this single, representative level of dispersion is constant throughout the meta-analysis~\citep[e.g.][]{higgins2009re, williams2021putting}.

A central challenge in real-world meta-analysis of promotional marketing experiments is predicting future experimental effects from accumulated past results. 
This challenge is complicated by datasets often involving multiple experiments with time-varying properties, leading to high heteroscedasticity and temporal heterogeneity. 
For example, in marketing, promotional experiments on digital platforms by retailers and media providers often exhibit such heterogeneity because they are conducted with diverse participants and media at different times. 
These conditions result in varying experimental means and variances, even among similar promotions. 
Consequently, predicting future experimental effects from these complex accumulated datasets is crucial for designing effective future promotional strategies. 
Since promotional effects can exhibit seasonality or other periodic patterns, capturing these periodic structures within a predictive framework can be particularly effective for accurate forecasting. 
Standard random-effects meta-analyses, however, typically assume homoscedasticity and neglect temporal heterogeneity, potentially inadequately addressing these complexities and thus limiting their predictive performance. 
Therefore, it is necessary to appropriately model high heteroscedasticity and complex time trends, including periodic components, to enhance understanding and predictive performance in such contexts.

Several approaches have been proposed to address high heteroscedasticity including between-study variances and complex time trends. 
\citet{thompson2020group} introduced weakly informative priors for subgroup-specific standard deviations. 
\citet{bowater2013heterogeneity} suggested making variance proportional to experiment-specific variables, such as participant numbers. 
\citet{williams2021putting} developed a Bayesian fixed-effect model using experiment-related variables for each experiment's variance. 
Nevertheless, these methods can lead to unreliable estimates due to large standard errors in small-scale experiments with limited sample sizes.
In meta-analysis literature addressing time effects, some methods compare time-varying functions across experiments using dose-response and time-response models~\citep{mawdsley2016model, pedder2019modelling}. 
Others treat experiment timing as fixed effects~\citep{miller2020empirical, salanti2009standard, vanEnst2015small}. 
However, the former does not account for treatment effects changing over time, limiting its applicability in scenarios with time-varying effects among experiments. 
The latter uses discrete dummy variables for time effects, leading to unstable estimates when time effects vary continuously.
Consequently, these approaches can be unsuitable for high heteroscedasticity and temporal heterogeneity.

To solve the aforementioned problems, we propose a new Bayesian estimation method called \textit{Shrinkage estimation with Time-trend and Random-Effects for meta-Analysis of Means and variance}~(STREAM). 
STREAM comprises two components: 
First, shrinkage estimation of means and variances through a hierarchical Bayesian approach that applies ``borrowing strength''.
In this component, while estimates of variances are produced, the principal motivation for robustly modeling these variances is to improve the performance and reliability of the mean effect estimation.
Second, estimation of time effects using the Gaussian process to flexibly model continuous-time trends in datasets.

We review related work on the two main components of our proposed method.
Modeling estimated variance is critical in small area estimation using random-effects \citep[e.g.][]{sugasawa2020small}. 
Notably, \citet{sugasawa2017bayesian} and \citet{gene2009empirical} employed random-effects models for both means and variance to achieve stable variance estimation under small sample sizes, resulting in accurate mean estimation. 
Such approaches are partially adopted in meta-analysis \citep[e.g.][]{wang2021variance}.
Observing parallels between small area estimation with area-level data and meta-analysis with experiment-level data, we develop the method to improve effect estimation in meta-analysis by shrinkage estimation of both means and variances. 
We also address temporal heterogeneity between experiments, an aspect less commonly considered in small area estimation.
\citet{roberts2013gaussian} introduces the usefulness of Gaussian processes in time series analysis. 
This approach is particularly beneficial when specific domain knowledge about the target time series is lacking but general knowledge exists, such as knowing that functions are smooth, continuous, or exhibit periodic variation.
In meta-analysis literature, \citet{leblanc2022time} investigated treatment resistance in resistant bacteria using Bayesian network meta-analysis, addressing temporal heterogeneity with a Gaussian process to model time trends. 
However, this approach neglected between-study heterogeneity by focusing on homogeneous experiments. 
We deal with substantial between-study heterogeneity by introducing random effects for both mean and variance.

The remainder of this paper is organized as follows. 
Section~\ref{sec:method} proposes a new Bayesian model for shrinkage estimation of experimental means and variances by incorporating hierarchical structures and accounting for time effects via a Gaussian process in meta-analysis. 
Section~\ref{sec:simulation} demonstrates the effectiveness of the proposed method through simulation studies. 
Section \ref{sec:app} illustrates its practical advantages using a real-world dataset, and Section~\ref{sec:conclusion} concludes the discussions.

\section{Gaussian Shrinkage Hierarchical Meta-analysis}
\label{sec:method}
\subsection{Setup and observed data}
We formulate the problem of meta-analysis for estimating experimental effects from a collection of $m$ individual experiments. 
For each experiment $i$~($i = 1, \dots, m$), let $\theta_i$ denote the true underlying effect or parameter of interest.
The observed data typically consist of $y_i$, which is an estimate of $\theta_i$, and an associated measure of its precision.
Specifically, $\sigma_i^2$ is defined as the true unknown sampling variance of the estimate $y_i$.
The data available from each experiment $i$, based on a sample of size $n_i$, includes $y_i$ and $S_i^2$, where $S_i^2$ is an observed estimate of this true sampling variance $\sigma_i^2$.

To build intuition, consider a common scenario where $y_i$ is the sample mean of $n_i$ independent observations drawn from a population with true mean $\theta_i$ and true variance $\psi_i^2$~(representing the variance of an individual observation in study $i$).
In this case, the true sampling variance of $y_i$~(which we denote as $\sigma_i^2$ throughout this paper) is given by $\sigma_i^2 = \psi_i^2 / n_i$, and this $\sigma_i^2$ is of order $n_i^{-1}$.
The observed $S_i^2$ would then be an estimate of this $\sigma_i^2$, typically calculated as $s_i^2 / n_i$, where $s_i^2$ is the sample variance of the $n_i$ observations (an estimate of $\psi_i^2$).

In conventional meta-analysis, the observed estimates $S_i^2$ are often treated as if they were the true sampling variances $\sigma_i^2$ and are used as fixed, known weights (typically the inverse of these variances) in pooling procedures.
Some simpler approaches might even disregard individual $S_i^2$ values if they assume a common variance across studies, though this is less common when study-specific variance estimates are available.
In contrast, our proposed model explicitly acknowledges $S_i^2$ as an estimate of an unknown $\sigma_i^2$ and models the relationship and uncertainty, as detailed in Equation~(\ref{eq:framework}). 

As auxiliary information, let $\bm{x}_i$ be a $q$-dimensional vector representing the characteristics of each experiment, and $\bm{z}_{ai}$ and $\bm{z}_{bi}$ be $J$-dimensional and $K$-dimensional dummy variables, respectively, indicating the specific groups to which experiment $i$ belongs.
For example, when focusing on company promotions, $\bm{z}_{ai}$ can represent the company, and $\bm{z}_{bi}$ can represent the type of industry. 
Although we consider two dummy variables as in light of the setting in data analysis in Section~\ref{sec:app}, extensions to handling more than two dummy variables are straightforward. 
Further, we assume that time information $t_i \in \{ t_1, \ldots, t_L \}$ indicating the time point of each experiment (e.g., the date it was conducted) is available.
Accordingly, let $\bm{z}_{ci}$ denote an $L$-dimensional dummy variable representing which specific time point from the set $\{ t_1, \ldots, t_L \}$ experiment $i$ corresponds to.

For the observed data $(y_i, S_i^2)$ from each experiment $i$, we assume the following distributions:
\begin{align}
\label{eq:framework}
\begin{split}
y_i \mid \theta_i, \sigma_i^2 &\sim N(\theta_i, \sigma_i^2), \\
S_i^2 \mid \sigma_i^2 &\sim \Gamma\left(\frac{n_i - 1}{2}, \frac{n_i - 1}{2\sigma_i^2}\right).
\end{split}
\end{align}
Here, $\Gamma(\alpha, \beta)$ denotes the gamma distribution with a probability density function proportional to $x^{\alpha - 1}\exp(-\beta x)$ for $x > 0$.
The distributional assumption for $S_i^2$ in (\ref{eq:framework}) is standard and implies that $(n_i - 1)S_i^2 / \sigma_i^2$ follows a $\chi^2$-distribution with $n_i-1$ degrees of freedom.
These assumptions for $y_i$ and $S_i^2$ are generally considered reasonable, particularly when the sample size $n_i$ for each study is not excessively small.
Furthermore, we assume that $y_i$ and $S_i^2$ are conditionally independent given their respective underlying parameters $\theta_i$ and $\sigma_i^2$.

\subsection{Models for heterogeneous means and variances}
To estimate $\theta_i$ and $\sigma_i^2$, we employ models with random effects estimated through a hierarchical Bayesian approach~\citep{lindley1972bayes}.
In particular, we model $\theta_i$ and $\sigma_i^2$ with effects of the auxiliary information and unobserved time effects by using a Gaussian process~\citep{brahim2004gaussian, williams2006gaussian}.

First, we express $\theta_i$ as follows.
\begin{align}
\label{eq:theta}
\begin{split}
& \theta_i = \alpha_{\theta} + \bm{\theta}_{a}^{\top} \bm{z}_{ai} + \bm{\theta}_{b}^{\top} \bm{z}_{bi} + \bm{\theta}_{c}^{\top} \bm{z}_{ci} + \bm{\beta}_{\theta}^{\top} \bm{x}_i , \\
& \theta_{aj} \sim N(0, \tau_a^2),  \ \ j = 1, \ldots, J, \ \ \ \theta_{bk} \sim N(0, \tau_b^2),  \ \ k = 1, \ldots, K, \\
& \bm{\theta}_{c} \sim N(0_L, C(\sigma_p^2, l_p)), \\
\end{split}
\end{align}
where $\alpha_{\theta}$ represents the baseline value common to each experiment, $\bm{\theta}_a$ and $\bm{\theta}_b$ denote the random effects for the groups to which each experiment belongs, expressed as $J$-dimensional and $K$-dimensional vectors, respectively.
Moreover, $\bm{\theta}_c$ represents the time effect.
$\bm{\beta}_{\theta}$ represents the regression coefficients for the experiment characteristics $\bm{x}_i$.
All elements of $\bm{\theta}_a = (\theta_{a1}, \ldots, \theta_{aJ})$ and $\bm{\theta}_b = (\theta_{b1}, \ldots, \theta_{bK})$ are assumed to be independent, and $\alpha$, $\bm{\beta}_{\theta}$, $\tau_a^2$ and $\tau_b^2$ are unknown parameters.
Here $C(\sigma_p^2, l_p)$ is the $L\times L$ variance-covariance matrix induced by the Gaussian process assumption on $\bm{\theta}_c$ as a function of $t_i$, and its $(l,l')$-element is $K(t_l, t_{l'}; \sigma_p^2, l_p)$ for some kernel function with parameters $\sigma_p^2$ and $l_p$. 
The kernel function, which determines the covariance structures of the Gaussian process, has various options. 
Since the data we will consider in our application contains time information as date, we use the following periodic covariance kernel \citep{roberts2013gaussian, williams2006gaussian} to take account of annual periodicity:
\begin{align*}
\begin{split}
& K(t, t'; \sigma_p^2, l_p) = \sigma_p^2 \exp{\left(- \frac{2\sin^2{(\pi |t - t'| / p_e)}}{l_p^2} \right)}, \\
\end{split}
\end{align*}
where $p_e$ is the length of period and $l_p$ and $\sigma_p$ are unknown parameters to characterize the kernel function. 

Regarding the variance parameter $\sigma_i^2$, we consider the following model: 
\begin{align}
\label{eq:sigma-squared}
\begin{split}
& \log \sigma_i^2 \sim N( \alpha_{\sigma} + \bm{\delta}_a^{\top} \bm{z}_{ai} + \bm{\delta}_b^{\top} \bm{z}_{bi} + \bm{\beta}_{\sigma}^{\top}\bm{x_i} - \log n_i,  \tau_{\sigma}^2), \\
& \delta_{aj} \sim N(0, \tau_{c}^2),  \ \ j = 1, \ldots, J, \ \ \ \delta_{bk} \sim N(0, \tau_{d}^2),  \ \ k = 1, \ldots, K, \\
\end{split}
\end{align}
where $\alpha_{\sigma}$ represents the baseline value common to each experiment, $\bm{\delta}_a$ and $\bm{\delta}_b$ denote the random effects for the groups to which each experiment belongs, expressed as $J$-dimensional and $K$-dimensional vectors, respectively, and $\bm{\beta}_{\sigma}$ represents the regression coefficients for the experiment characteristics $\bm{x}_i$.
Further, $\alpha_{\sigma}$, $\bm{\beta}_{\sigma}$, $\tau_{\sigma}^2$, $\tau_{c}^2$ and $\tau_{d}^2$ are unknown parameters.
Note that $-\log n_i$ in the mean term of (\ref{eq:sigma-squared}) corresponds to an offset term to ensure that $E[\sigma_i^2]=O(n_i^{-1})$.
The proposed model consists of (\ref{eq:framework}), (\ref{eq:theta}) and (\ref{eq:sigma-squared}), and its advantages can be summarized as follows: 
\begin{itemize}
\item[-]
Shrinkage estimation for both the mean and variance allows us to account for the dataset's heteroscedasticity and achieve robust mean estimates when predicting $\theta_i$ for unknown experiments.

\item[-]
The time trend modeled by a Gaussian process enables us to predict future effects of experiments.   
\end{itemize}

\subsection{Posterior inference and predicting future effects}
The proposed model \eqref{eq:framework}, \eqref{eq:theta}, and \eqref{eq:sigma-squared} includes several unknown parameters. 
By assigning prior distributions to them, we consider Bayesian inference on the model parameters and the random effects (latent variables). 
As the default choice, we adopt relatively standard prior distributions as follows.
\begin{align}
\label{eq:prior}
\begin{split}
& \alpha_{\theta} \sim N(m_{\alpha \theta}, s_{\alpha \theta}^2), \ \
\alpha_{\sigma} \sim N(m_{\alpha \sigma}, s_{\alpha \sigma}^2), \ \ 
\bm{\beta}_{\theta} \sim N(m_{\beta \theta}, s_{\beta \theta}^2), \\
& \bm{\beta}_{\sigma} \sim N(m_{\beta \sigma}, s_{\beta \sigma}^2), \ \ 
\tau_a^2\sim {\rm HC}(\eta_a), \ \   
\tau_b^2\sim {\rm HC}(\eta_b), \ \
\tau_c^2 \sim {\rm HC}(\eta_c), \\
&\tau_d^2 \sim {\rm HC}(\eta_d), \ \
\tau_{\sigma}^2 \sim {\rm HC} (\eta_e), \ \ 
\sigma_p^2\sim {\rm HC}(\eta_{\sigma}), \ \
l_p \sim {\rm HC}(\eta_l), \ \ 
\end{split}
\end{align}
where ${\rm HC}(\eta)$ denotes the half-Cauchy prior \citep[e.g.][]{gelman2006prior} with scale parameter $\eta$. 
The values in the above prior distributions (hyperparameters) are user-specified, and specific choices will be given in Sections~\ref{sec:simulation} and \ref{sec:app}.
We note that the log-normal model for $\sigma_i^2$ in (\ref{eq:sigma-squared}) is not conjugate under the sampling model (\ref{eq:framework}), which may complicate the posterior inference.
To efficiently generate posterior samples, we rely on the Markov chain Monte Carlo (MCMC) algorithm, implemented in the probabilistic programming \texttt{stan}~\citep{carpenter2017stan}.
Stan is a probabilistic programming language widely used for Bayesian modeling, which leverages the Hamiltonian Monte Carlo algorithm~\citep{duane1987hybrid} to efficiently sample from complex probability distributions.

We calculate the predictive distribution of the estimates $\tilde{\bm{y}} = (\tilde{y}_1, \ldots, \tilde{y}_{m_{\text{te}}})$ for $m_{\text{te}}$ new future experiments with specific time point from the set $\{t^{\prime}_{1}, \ldots, t^{\prime}_{L^{\prime}} \}$.
Specifically, we use $m_{\text{tr}}$ experiments as observational data and the posterior samples of each parameter generated by MCMC.
In constructing the predictive distribution of $\tilde{\bm{y}}$, we use the posterior samples generated by MCMC for various parameters in equations \eqref{eq:framework}, \eqref{eq:theta}, and \eqref{eq:sigma-squared}. 
We provide additional explanation since constructing the predictive distribution for $\bm{\theta}_c$ differs from other parameters.
Our goal is to construct the predictive distribution of the parameters $\tilde{\bm{\theta}}_{c} = (\tilde{\theta}_{c1}, \ldots, \tilde{\theta}_{c L^{\prime}})$ associated with the future experiments $\tilde{\bm{z}}_{c} = (\tilde{\bm{z}}_{c1}, \ldots, \tilde{\bm{z}}_{c L^{\prime}})$, using the posterior samples.
To achieve this, we aim to obtain the predictive distribution $p(\tilde{\bm{\theta}}_{c} | D_{\text{tr}}, \tilde{\bm{z}}_{c})$.
Here, $D_{\text{tr}} = (\bm{z}_{c}, \bm{\theta}_{c})$, where $\bm{z}_{c} = (\bm{z}_{c1}, \ldots, \bm{z}_{c L})$ and $\bm{\theta}_{c} = (\theta_{c1}, \ldots, \theta_{c L})$.
Since the joint distribution $p(\tilde{\bm{\theta}}_{c}, \bm{\theta}_{c} | \bm{z}_{c}, \tilde{\bm{z}}_{c})$ is given by the Gaussian process assumption in equation~\eqref{eq:theta}, it is formulated as the following multivariate Gaussian distribution.
\begin{align}
\begin{split}
p(\bm{\theta}_{c}, \tilde{\bm{\theta}}_{c} | \bm{z}_{c}, \tilde{\bm{z}}_{c}) \sim N(0_{L + L^{\prime}}, C^{\prime}(\sigma^2_p, l_p))
\end{split}
\end{align}
where $C^{\prime}(\sigma_p^2, l_p)$ is the $(L + L^{\prime})\times (L + L^{\prime})$ the following variance-covariance matrix:
\begin{align}
\begin{split}
C'(\sigma^2_p, l_p) = 
\begin{pmatrix} 
\bm{\Phi}_{L L}                 & \bm{\Phi}_{L L^{\prime}} \\
\bm{\Phi}^{\top}_{L L^{\prime}} & \bm{\Phi}_{L^{\prime} L^{\prime}}
\end{pmatrix}
\end{split}
\end{align}
whose elements are given by
\begin{align}
\begin{split}
&\bm{\Phi}_{L L} = (K(t_i, t_j; \sigma^2_p, l_p)), \ \ 1 \leq i, j \leq L, \\
&\bm{\Phi}_{L L^{\prime}} = (K(t_i, t^{\prime}_j; \sigma^2_p, l_p)), \ \ 1 \leq i \leq L, \ \ 1 \leq j \leq L^{\prime} \\
&\bm{\Phi}_{L^{\prime} L^{\prime}} = (K(t^{\prime}_i, t^{\prime}_j; \sigma^2_p, l_p)), \ \ 1 \leq i, j \leq L^{\prime}, \\
\end{split}
\end{align}
By using the property of conditional multivariate Gaussian~\citep[e.g.][]{williams2006gaussian}, we can obtain the predictive distribution $p(\tilde{\bm{\theta}}_{c} | D_{\text{tr}}, \tilde{\bm{z}}_{c})$ as follows.
\begin{align}
\begin{split}
p(\tilde{\bm{\theta}}_c | D_{tr}, \tilde{\bm{z}}_c) 
= N(\bm{\Phi}^{\top}_{L L^{\prime}} \bm{\Phi}^{-1}_{L L} \bm{\theta}_c, \bm{\Phi}_{L^{\prime} L^{\prime}} - \bm{\Phi}^{\top}_{L L^{\prime}} \bm{\Phi}^{-1}_{L L} \bm{\Phi}_{L L^{\prime}}).
\end{split}
\end{align}
Then, using the posterior samples of $\alpha_\theta, \bm{\theta}_{a}, \bm{\theta}_{b}, \tilde{\bm{\theta}}_{c}$ and $ \bm{\beta}_{\theta}$, the posterior samples of the future effect $\tilde{\theta}_j$ can be generated as 
$$
\tilde{\theta}_j = \alpha_{\theta} + \bm{\theta}_{a}^{\top} \tilde{\bm{z}}_{aj} + \bm{\theta}_{b}^{\top} \tilde{\bm{z}}_{bj} + \tilde{\bm{\theta}}_{c}^{\top} \tilde{\bm{z}}_{cj} + \bm{\beta}_{\theta}^{\top} \tilde{\bm{x}}_j, \ \ \ \ j=1,\ldots,m_{\text{te}},
$$
where $\tilde{\bm{z}}_{aj}$, $\tilde{\bm{z}}_{bj}$ and $\tilde{\bm{x}}_j$ are covariate vectors for the test data. 
This parameter can be regarded as the true effect in the future treatment. 
Furthermore, given covariates $\tilde{\bm{z}}_{aj}$, $\tilde{\bm{z}}_{bj}$, $\tilde{\bm{x}}_j$ and the sample size $n_j$, the posterior samples of the future variance $\tilde{\sigma}_j^2$ can be generated as 
$$ 
\log \tilde{\sigma}_j^2 \sim N( \alpha_{\sigma} + \bm{\delta}_a^{\top} \tilde{\bm{z}}_{aj} + \bm{\delta}_b^{\top} \tilde{\bm{z}}_{bj} + \bm{\beta}_{\sigma}^{\top} \tilde{\bm{x}}_j - \log n_j,  \tau_{\sigma}^2),
$$
so that the posterior predictive distribution of the observed treatment effect $\tilde{y}_j$ can be generated from $N(\tilde{\theta}_j ,\tilde{\sigma}_j^2)$.

\section{Simulation Studies}
\label{sec:simulation}
We compare the performance of the proposed method with baseline methods through simulation studies. 
The $m$ experimental datasets subject to meta-analysis are represented by $(y_i, S_i^2, n_i)$ for $i=1, \ldots, m$, where $y_i$ is the estimates of the mean effect of interest $\theta_i$ in experiment $i$, $S_i^2$ is the variance of the $y_i$ for experiment $i$, and $n_i$ is the sample size of experiment $i$. 
In the simulation experiments, to generate the final dataset $(y_i, S_i^2, n_i)$, we perform data generation using the following steps.

First, we generate the observations $y_{ij}$ for experiment $i$ from the following:
\begin{align}
\begin{split}
&y_{ij} \sim N(\theta_i, n_i \sigma_i^2), \ \ \ i = 1, \ldots, m,\  j = 1, \ldots, n_i, \\
\end{split}
\end{align}
Using $y_{ij}$ and $n_i$, we calculate the mean $y_i$ and variance $S_i^2$ for experiment $i$ as follows.
\begin{align}
\begin{split}
&y_i = \frac{1}{n_i} \sum_{j=1}^{n_i} y_{ij}, \quad S_i^2 = \frac{1}{n_i(n_i - 1)} \sum_{j=1}^{n_i} (y_{ij} - y_i)^2. 
\end{split}
\end{align}

We then generate the parameter of interest $\theta_i$ using the model below, which accounts for random and time effects.
\begin{align}
\label{eq:sim-theta}
\begin{split}
&\theta_i = \bm{\theta}_{a}^{\top} \bm{z}_{ai} + \bm{\theta}_{b}^{\top} \bm{z}_{bi} + a_{1} \sin\left(2\pi \frac{t_i}{12}\right) + b_{1} \cos\left(2\pi \frac{t_i}{12}\right) + c_{1} \frac{t_i}{12} + 0.5 x_i, \\
&\theta_{aj}, \theta_{bk} \sim N(\mu_{\theta}, \sigma_{\theta}^2), \ \ \ \mu_{\theta} \sim N(3, 2^2), \ \ \ \sigma_{\theta}^2 \sim N_{[0, \infty)}(0, d_{1}^2), \\
\end{split}
\end{align}

We also generate $\sigma_i^2$ using the following model that accounts for random effects.
\begin{align}
\begin{split}
&\sigma_i^2 \sim IG(2, \exp(\bm{\delta}_{a}^{\top} \bm{z}_{ai} + \bm{\delta}_{b}^{\top} \bm{z}_{bi} + 0.1 x_i)), \\
&\delta_{aj}, \delta_{bk} \sim N(\mu_{\sigma}, \sigma_{\sigma}^2), \ \ \ \mu_{\sigma} \sim N(1, 0.1^2), \ \ \ \sigma_{\sigma}^2 \sim N_{[0, \infty)} (0, d_{2}^2).
\end{split}
\end{align}
We generate the covariate $x_i$ from a uniform distribution between 1 and 10. 
The time information $t_i$ takes random values from ${1, \ldots, 24}$, and the sample size $n_i$ takes random values from ${100, \ldots, 10000}$. 
In this study, we set $J=30$, $K=6$, and $m=80$. We split each simulation dataset into training and test sets in an 8:2 ratio.
The test set is taken in order from the largest values of $t_i$, so it is the future promotion compared to the training set. 

In this study, we conduct simulations based on four scenarios that combine the presence or absence of time effects and the magnitude of between-study heterogeneity. 
The presence of time effects is determined by the existence of $a_1$, $b_1$, and $c_1$ in Equation (\ref{eq:sim-theta}), and the magnitude of between-study heterogeneity is determined by the values of $d_1$ and $d_2$ in $\sigma_{\theta}^2$ and $\sigma_{\sigma}^2$. 
We perform simulations based on the following four scenarios: (i) temporal heterogeneity present and high between-study heterogeneity $(a_1, b_1, c_1, d_1, d_2) = (1, 1, 1, 10, 2)$; (ii) temporal heterogeneity present and low between-study heterogeneity $(a_1, b_1, c_1, d_1, d_2) = (1, 1, 1, 5, 0.5)$; (iii) temporal heterogeneity absent and high between-study heterogeneity $(a_1, b_1, c_1, d_1, d_2) = (0, 0, 0, 10, 2)$; and (iv) temporal heterogeneity absent and low between-study heterogeneity $(a_1, b_1, c_1, d_1, d_2) = (0, 0, 0, 5, 0.5)$.

For each simulation scenario, we generate 150 simulation datasets and estimate the experimental effect $\theta_i$ using seven baseline methods and the proposed STREAM:
(a) Fixed Effects~(FE) model,
(b) FE with monthly dummy~(FE-M) model,
(c) FE with monthly dummy and variance modeling~(FE-MV) model,
(d) Random Effects~(RE) model,
(e) RE with monthly dummy~(RE-M) model,
(f) RE with monthly dummy and variance modeling~(RE-MV) model,
(g) RE-Gaussian-Process~(RE-GP) model, and
(h) STREAM.
The FE model, corresponding to the fixed effects model in general meta-analysis~\citep{borenstein2010basic}, models $\bm{\theta}_a$ and $\bm{\theta}_b$ as fixed effects in equation \eqref{eq:theta} and does not account for the time effect $\bm{\theta}_c$, using an uninformative prior distribution $N(0, 1000)$ for $\bm{\theta}_a$ and $\bm{\theta}_b$.
The FE-M model extends the FE model by including the time effect $\bm{\theta}_c$ as a fixed effect, with an uninformative prior $N(0, 1000)$ for $\bm{\theta}_c$. 
The FE-MV model further incorporates fixed effects for $\bm{\delta}_a$ and $\bm{\delta}_b$ in equation \eqref{eq:sigma-squared}, also using an uninformative prior $N(0, 1000)$. 
This model corresponds to existing methods by \citet{bowater2013heterogeneity} and \citet{williams2021putting}.
The RE model, corresponding to hierarchical Bayesian random effects modeling in general meta-analysis~\citep{babapulle2004hierarchical, borenstein2010basic}, models $\bm{\theta}_a$ and $\bm{\theta}_b$ as random effects in equation \eqref{eq:theta} without considering the time effect $\bm{\theta}_c$.
The RE-M model extends the RE model by modeling the time effect $\bm{\theta}_c$ as a fixed effect with an uninformative prior $N(0, 1000)$. 
The RE-MV model further incorporates random effects for $\bm{\delta}_a$ and $\bm{\delta}_b$ in equation \eqref{eq:sigma-squared}.
The RE-GP model models $\bm{\theta}_a$ and $\bm{\theta}_b$ as random effects in equation \eqref{eq:theta} and represents the time effect $\bm{\theta}_c$ using a Gaussian process, corresponding to the method in \citet{leblanc2022time}.

These baseline methods adopt distinct approaches for incorporating and modeling $S_i^2$.
Broadly, they are classified based on whether they explicitly model variance heterogeneity.
For the first group of baseline methods–FE, RE, FE-M, RE-M, and RE-GP–which do not explicitly model variance heterogeneity, the observed variance estimate for each study $i$, $S_i^2$, is treated as a known and fixed value representing the sampling variance $\sigma_i^2$ for that study. 
Consequently, each study's observed effect $y_i$ is assumed to follow a normal distribution with mean $\theta_i$ and this fixed variance $S_i^2$. 
Subsequently, when predicting the effect of new experiments for which individual variances are unknown, these baseline methods use a single, fixed variance for all such predictions; this fixed variance is calculated as the average of the observed $S_i^2$ values from the training dataset.

In contrast, the FE-MV and RE-MV explicitly model the observed sample variance $S_i^2$ conditional on the true sampling variance $\sigma_i^2$ using a Gamma distribution, as specified in Equation~\ref{eq:framework}.
This modeling strategy for $S_i^2 \mid \sigma_i^2$ is equivalent to the standard statistical assumption that $(n_i-1)S_i^2 / \sigma_i^2$, follows a Chi-squared distribution.
Such an approach to modeling sample variances via a Gamma distribution has precedent in the meta-analysis literature~\citep[e.g.][]{chowdhry2016meta, dakin2011mixed}.

Bayesian estimation is performed for these models by setting uninformative priors for each relevant parameter.
\begin{align}
\begin{split}
& m_{\alpha \theta}=m_{\alpha \sigma}=0, \ \ \  m_{\beta \theta}=m_{\beta \sigma}=0_{q}, \\
& s_{\alpha \theta}=s_{\alpha \sigma}=1000, \ \ \ s_{\beta \theta}=s_{\beta \sigma}=1000_{q}, \\
& \eta_a=\eta_b=\eta_c=\eta_d=\eta_e=\eta_l=\eta_{\sigma}=2.5.
\end{split}
\end{align}
We set $p_e=12$ for the periodic covariance kernel of the Gaussian process.
For each simulation and model, we executed the MCMC algorithm with four chains, each comprising 2,000 warm-ups and 8,000 sampling iterations.
The point estimates of the model parameters are calculated using the median of the posterior samples for each model.

We evaluate the models in terms of their prediction performance in point and interval estimation. 
The point estimation performance of the models is compared based on the mean average percentage error (MAPE) given by
\begin{equation}
 \text{MAPE} = \frac{100}{n} \sum_{i=1}^{n} \Big\|\frac{\theta_i - \hat{y}_i}{\theta_i}\Big\|,
\end{equation}
and the scaled mean squared error~(scaled MSE) given by 
\begin{equation}
\text{scaled MSE} = \frac{1}{n} \sum_{i=1}^{n} \frac{(\theta_i - \hat{y}_i)^2}{\theta_i^2},
\end{equation}

where $\hat{y}_i$ is the predicted treatment effect of the experiment $i$, and $n$ is the number of experiments to be predicted. 

We use the interval score~(IS)~\citep{gneiting2007strictly} to evaluate the interval estimation performance of the models.
Specifically, the IS for the $100(1 - \alpha)$\% prediction interval, with lower and upper endpoints that are the predictive quantiles at level $\alpha/2$ and $(1-\alpha)/2$, respectively, are defined below.
\begin{align}
\begin{split}
\text{IS} = \frac{1}{n} \sum_{i=1}^{n} \left\{(u_i -l_i) + \frac{2}{\alpha} (l_i - \theta_i) \mathbbm{1}(\theta_i < l_i) + \frac{2}{\alpha} (\theta_i - u_i) \mathbbm{1}(\theta_i > u_i) \right\},
\end{split}
\end{align}
where $l_i$ and $u_i$ are the lower and upper bounds of the $100(1 - \alpha)$\% highest posterior density intervals, respectively, and $\mathbbm{1}(\cdot)$ is the indicator function.
The IS is a measure designed to quantify the quality of uncertainty in the model predictions. It penalizes the model if the prediction interval it outputs is vast and penalizes the model if the observed values are outside the prediction interval it outputs.
In this paper, $\alpha$ is 0.05. 

Table~\ref{tab:simulation-w} presents the simulation results for scenarios (i) and (ii), which include temporal heterogeneity.
In simulation~(i), STREAM demonstrated the best performance in both point estimation~(MAPE and Scaled MSE) and interval estimation~(IS). 
In simulation~(ii), STREAM achieved the best performance in interval estimation~(IS), while FE-MV showed the best performance in point estimation~(MAPE and Scaled MSE). 
Although STREAM did not secure the top spot in point estimation, it ranked third overall in MAPE and Scaled MSE. 
\begin{table}[tb]
\centering
\caption{Simulation results including temporal heterogeneity. The boldface values indicate the best performance among all models, while underlined values represent the second-best results.}
\label{tab:simulation-w}
\begin{tabular}{ccrrr}
\hline
                      & Method          & \multicolumn{1}{c}{MAPE}           & \multicolumn{1}{c}{Scaled MSE}  & \multicolumn{1}{c}{IS}                 \\ \hline
\multirow{8}{*}{(i)}  & FE              & 152.8           & 1894.84          & 34.8                \\
                      & FE-M            & 65.4            & 263.62           & 23.5                \\
                      & FE-MV           & 87.5            & 839.62           & $>1000$              \\
                      & RE              & 78.1            & 158.80           & 35.1                \\
                      & RE-M            & 61.3            & 147.14           & 21.9                \\
                      & RE-MV           & \underline{36.9}& \underline{23.75}& \underline{13.8}    \\
                      & RE-GP           & 69.2            & 58.27            & 38.4                \\
                      & \textbf{STREAM}& \textbf{29.7}   & \textbf{8.12}     & \textbf{10.0}        \\ \hline
\multirow{8}{*}{(ii)} & FE              & 48.1            & 46.61            & 33.4                \\
                      & FE-M            & 28.5            & 23.57            & 18.4                \\
                      & FE-MV           & \textbf{21.1}   & \textbf{18.11}   & $>1000$             \\
                      & RE              & 36.1            & 23.57            & 23.4                \\
                      & RE-M            & 22.8            & \underline{21.22}& 8.2                 \\
                      & RE-MV           & \underline{21.6}& 21.51            & 9.0                 \\
                      & RE-GP           & 23.9            & 21.74            & \underline{6.1}     \\
                      & \textbf{STREAM}& 22.0            & 21.50             & \textbf{5.4}        \\ \hline 
\end{tabular}
\end{table}

Table~\ref{tab:simulation-wo} presents the simulation results for scenarios (iii) and (iv), which do not include temporal heterogeneity.
In simulation ~(iii), STREAM performed best in both point estimation~(MAPE and Scaled MSE) and interval estimation~(IS). 
In simulation~(iv), STREAM showed the highest performance in point estimation~(MAPE, Scaled MSE) and third in IS.

\begin{table}[tb]
\centering
\caption{Simulation results without temporal heterogeneity. The boldface value indicate the best performance among all models, while underlined values represent the second-best results.}
\label{tab:simulation-wo}
\begin{tabular}{ccrrr}
\hline
                      & Method          & \multicolumn{1}{c}{MAPE}            & \multicolumn{1}{c}{Scaled MSE}  & \multicolumn{1}{c}{IS}                 \\ \hline
\multirow{8}{*}{(iii)}& FE              & 49.9            & 67.10            & 19.5                \\
                      & FE-M            & 63.9            & 159.56           & 22.8                \\
                      & FE-MV           & 38.4            & 31.72            & $>1000$              \\
                      & RE              & 43.7            & 41.12            & 18.3                \\
                      & RE-M            & 52.1            & 74.31            & 21.6                \\
                      & RE-MV           & \underline{28.0}& \underline{5.21} & \underline{13.6}    \\
                      & RE-GP           & 429.3           & 32983.78         & 34.7                \\
                      & \textbf{STREAM} & \textbf{22.2}   & \textbf{4.36}    & \textbf{9.6}        \\ \hline
\multirow{8}{*}{(iv)} & FE              & 31.0            & 21.41            & 14.2                \\
                      & FE-M            & 30.0            & 11.61            & 17.8                \\
                      & FE-MV           & 19.6            & 5.76             & $>1000$              \\
                      & RE              & 16.6            & 4.13             & \textbf{4.2}        \\
                      & RE-M            & 19.4            & 3.83             & 8.0                 \\
                      & RE-MV           & 18.3            & \underline{3.50} & 9.0                 \\
                      & RE-GP           & \underline{16.4}& 3.92             & \underline{4.4}     \\
                      & \textbf{STREAM} & \textbf{15.9}   & \textbf{3.40}    & 5.1                 \\ \hline
\end{tabular}
\end{table}

From these four simulation results, we found that STREAM consistently outperforms other baseline methods in scenarios with high between-study heterogeneity (simulations~(i) and~(iii)), irrespective of temporal heterogeneity. 
Furthermore, when between-study heterogeneity is low (simulations~(ii) and~(iv)), STREAM maintains stable performance in both point and interval estimation.

In scenarios with high between-study heterogeneity, STREAM outperforms other baseline methods, with RE-MV being the second-best model. 
This demonstrates the effectiveness of Gaussian process modeling for more flexible time effect modeling compared to RE-MV. 
Conversely, RE-GP, which does not account for variance modeling, performs similarly to STREAM when between-study heterogeneity is low but significantly deteriorates when heterogeneity is high. 
This deterioration is likely due to RE-GP's inability to adequately model the data's variance structure. 
This causes the Gaussian process for time effects to overfit the data as it attempts to compensate for incorrect parameter structures, thereby reducing predictive performance.
These findings show the importance of our proposed method, STREAM, which addresses high between-study heterogeneity in meta-analysis by modeling both means and variances through a hierarchical Bayesian approach while flexibly representing time effects with a Gaussian process. 
In simulation ~(ii), despite its excellent performance in some aspects, the FE-MV showed the worst performance in IS across all simulations. 
This is likely because the parameters for unknown categories, $\bm{\theta}_a$, $\bm{\theta}_b$, $\bm{\delta}_a$, and $\bm{\delta}_b$, were not updated and were strongly influenced by non-informative priors. 
This result illustrates the importance of the hierarchical Bayesian approach, which shares information to achieve stable estimation even for unknown categories.

\section{Application to Promotion Effects in Marketing}\label{sec:app}
We apply the proposed model to meta-analysis in promotions implemented as marketing strategies.
Meta-analysis is used in promotions to predict promotion effectiveness using results from past promotions and improve corporate revenues by identifying and implementing promotions with high Return on Investment. 
Meta-analysis in marketing is common not only in e-commerce, where the accumulation and use of user data are actively conducted~\citep{browne2017works, miller2020empirical, zerbini2022drivers}, but also in retail businesses with offline stores~\citep{blut2018testing, liuthompkins2022what}. 
Practically, since numerous experiments are added daily and need to be frequently updated, the meta-analysis approach using summary statistics of existing experiments has a notable advantage in ease of implementation and scalability.

Our proposed method enables more accurate estimates when the experiments under analysis exhibit high between-study and temporal heterogeneity. By utilizing our proposed method, marketers can make highly confident decisions about which marketing strategies are most effective for implementing more profitable campaigns.

In this study, we use a real-world dataset of promotions aimed at sales promotion conducted by participating merchants within a coalition loyalty program~(a system that allows flexible acquisition and redemption of points among multiple participating stores)~\citep[e.g.][]{stourm2023cross}. 
This promotion dataset records, for each promotion, the participating merchant, business type, promotion start/end dates, number of subjects in the experiment (with equal numbers in treatment and control groups), estimated treatment effect, variance, and promotion duration. 
The data analyzed in this study comprises promotions conducted from April 2022 to December 2023. 
From this promotion data, we extract and analyze data from companies that conducted two or more promotions during the period. 
The dataset includes 51 promotions carried out by 16 participating merchants in two types of businesses: retail and restaurant.

\begin{figure}[tb]
\centering
\begin{minipage}[b]{0.49\textwidth}
\includegraphics[width=\textwidth]{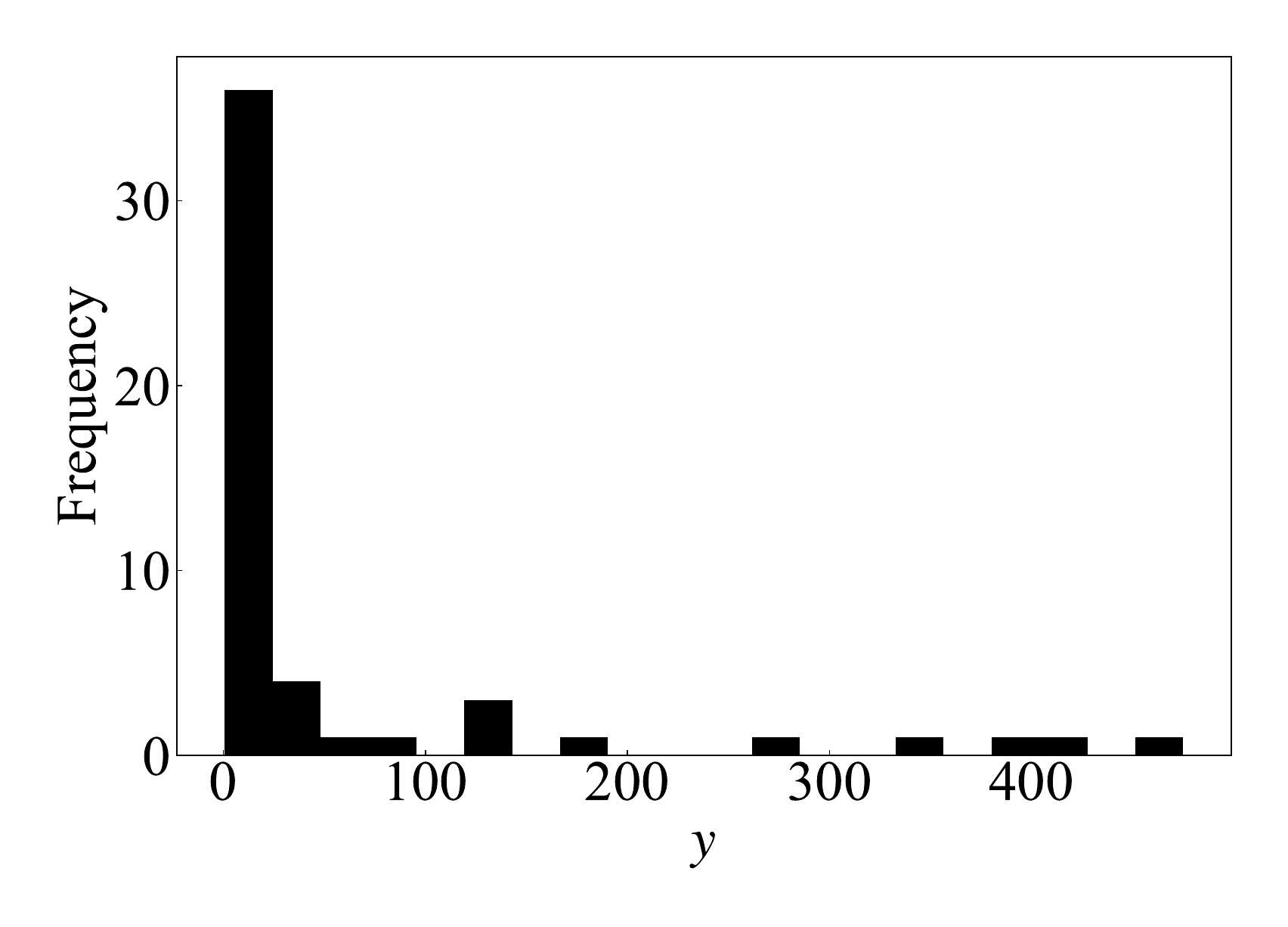}
\end{minipage}
\begin{minipage}[b]{0.49\textwidth}
\includegraphics[width=\textwidth]{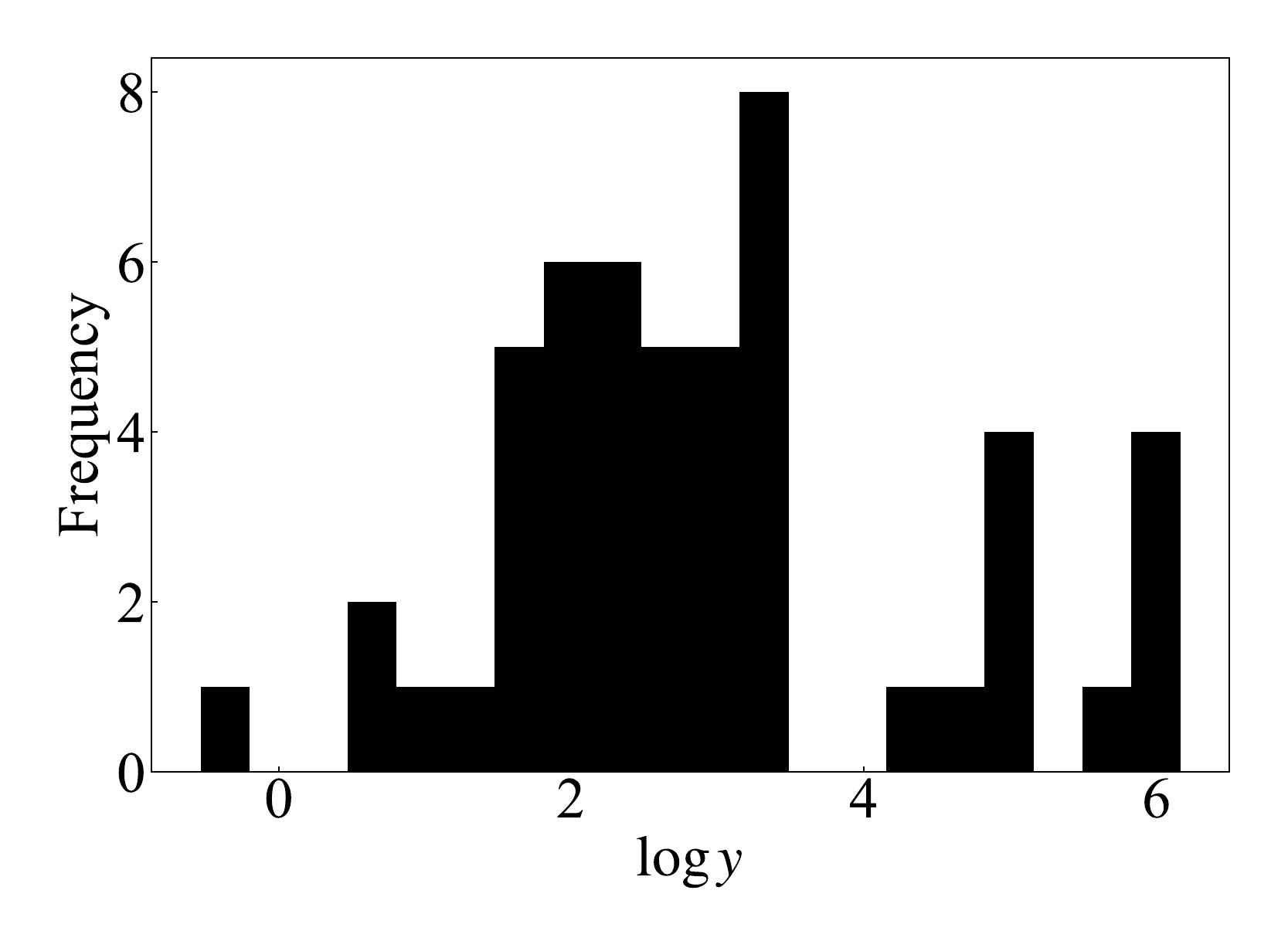}
\end{minipage}
\caption{Comparison of the distribution of the original variable $y$ and the log-transformed variable $\log y$. }
\label{fig:y}
\end{figure}

We set the promotion duration as the covariate $x_i$. 
Table~\ref{tab:summary-stat} presents the summary statistics for the sample size $n_i$, the estimated mean treatment effect $y_i$, variance $S_i^2$, and the covariate $x_i$ of the dataset.
\begin{table}[tb]
\centering
\caption{Summary statistics for the sample size $n_i$, the estimated mean treatment effect $y_i$, variance $S_i^2$, and the covariate $x_i$ of the dataset.}
\label{tab:summary-stat}
\begin{tabular}{crr}
\hline
                  & \multicolumn{1}{c}{Mean}    & \multicolumn{1}{c}{St. Dev.} \\ \hline
Sample size $n_i$ & 26573.0 & 31519.5\\
Mean $y_i$        & 61.7    & 115.0  \\
Variance $S_i^2$  & 22.4    & 76.7   \\
Covariance $x_i$  & 24.9    & 7.5    \\ \hline
\end{tabular}
\end{table}
This dataset consists of promotions conducted by multiple merchants at different time points, resulting in substantial between-study and temporal heterogeneity.
Figure \ref{fig:y} shows the distribution of the original variable $y$ and the log-transformed variable $\log y$.
As shown in Figure~\ref{fig:y}, this dataset's distribution of $y$ is asymmetric due to substantial between-study heterogeneity and outliers.
Also, the distribution of $\log y$ is more symmetric and reduces the effect of outliers, making it more suitable for linear modeling.
Therefore, the analysis uses the log-transformed variable $\log y$ as the response variable.
We apply the Delta method~\citep{oehlert1992note} to transform the variance associated with the log transformation of the estimator $y$. The variance of the log-transformed estimator $\log y$ can be approximated as $\text{Var} (\log y) \thickapprox \text{Var} (y) / y^2$.

We split the dataset into training and test sets in an 8:2 ratio.
The test set is taken in order from the largest values of $t_i$, so it is the future promotion compared to the training set.
We executed the MCMC algorithm with four chains comprising 2,000 warm-ups and 8,000 sampling iterations.
Using this real-world dataset, we evaluate seven baseline methods and the proposed method: 
(a) FE model,
(b) FE-M model,
(c) FE-MV model,
(d) RE model,
(e) RE-M model,
(f) RE-MV model,
(g) RE-GP model, and
(h) STREAM.

Table \ref{tab:application} shows the results of the application to the promotion effect data.
\begin{table}[tbp]
\caption{Performance metrics from the application to the promotion effect dataset. The boldface values indicate the best performance among all models, while underlined values represent the second-best results.}
\label{tab:application}
\centering
\begin{tabular}{crrr}
\hline
Method & \multicolumn{1}{c}{MAPE} & \multicolumn{1}{c}{Scaled MSE}  & \multicolumn{1}{c}{IS} \\ \hline
FE               & 77.8                     & 0.96                & 531.6 \\
FE-M             & 70.2                     & 0.58                & 588.1 \\
FE-MV            & 68.9                     & 0.54                & $>1000$ 
\\
RE               & 108.8                    & 2.64                & 530.1\\
RE-M             & 67.4                     & 0.49                & 603.1\\
RE-MV            & \underline{50.4}         & \underline{0.32}    & \underline{473.7} \\
RE-GP            & 135.5                    & 4.04                & 547.8\\
\textbf{STREAM}  & \textbf{48.1}            & \textbf{0.29}       & \textbf{464.5} \\ \hline
\end{tabular}
\end{table}
From these results, the proposed method STREAM demonstrated the best performance in MAPE, Scaled MSE, and IS compared to other baseline methods. 
Comparisons between the FE and FE-M and the RE and RE-M revealed that accounting for time effects improves predictive performance. 
The RE-MV model exhibited the second-best performance, consistent with the trends observed in simulation~(i) in Table~\ref{tab:simulation-w}. 
Furthermore, the observed deterioration in RE-GP's performance aligns with the results of simulation~(i), suggesting that the real dataset exhibits significant between-study heterogeneity and temporal heterogeneity. 
Simulation studies and the real dataset confirmed that STREAM exhibits superior performance on datasets with high between-study and temporal heterogeneity.

To verify the convergence of each model, we calculated the Gelman-Rubin~(GR) statistic $\hat{R}$~\citep{gelman1992inference} for each model. 
Table~\ref{tab:R-hat} presents each model's median and maximum values of the GR statistic $\hat{R}$.
\begin{table}[tbp]
\label{tab:R-hat}
\centering
\caption{Median and maximum values of Gelman-Rubin~(GR) statistic $\hat{R}$. The boldface values indicate $\hat{R} < 1.1$, satisfying the convergence criterion for MCMC.}
\begin{tabular}{crr}
\hline
Method           & \multicolumn{1}{c}{Median $\hat{R}$} & \multicolumn{1}{c}{Max $\hat{R}$} \\ \hline
FE               & 2.2                                  & 3.8           \\
FE-M             & 2.7                                  & 4.0           \\
FE-MV            & 1.4                                  & 2.7           \\
RE               & 2.1                                  & 3.5           \\
RE-M             & 2.8                                  & 4.3           \\
RE-MV            & \textbf{1.0}                         & \textbf{1.0}  \\
RE-GP            & 3.1                                  & 4.1           \\
\textbf{STREAM} & \textbf{1.0}                         & \textbf{1.0}  \\ \hline
\end{tabular}
\end{table}
According to \citet{gelman2013bayesian}, a GR statistic $\hat{R}$ of 1.1 or below indicates convergence. 
Under these diagnostic criteria, the baseline methods FE, FE-M, FE-MV, RE, RE-M, and RE-GP exhibited poor MCMC convergence. In contrast, RE-MV and STREAM were confirmed to have achieved convergence. This is likely because RE-MV and the proposed STREAM effectively capture the data characteristics through hierarchical Bayesian approaches and time effect modeling, resulting in stable parameter sampling.

\begin{figure}[tbp]
\centering
\includegraphics[width=\textwidth]{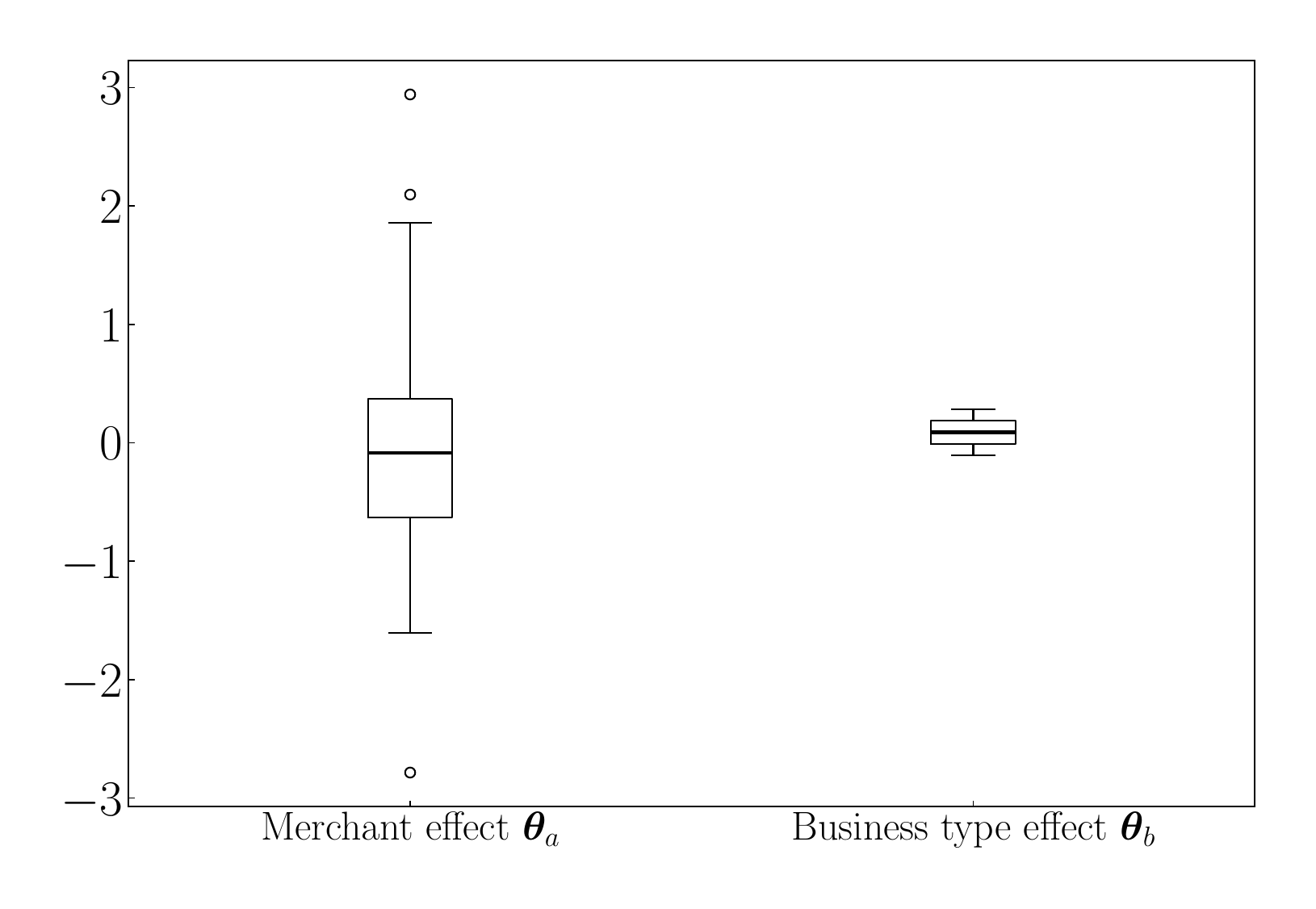}
\caption{Estimated merchant and business type effects.}
\label{fig:random-effects}
\end{figure}

\begin{figure}[tbp]
\centering
\includegraphics[width=\textwidth]{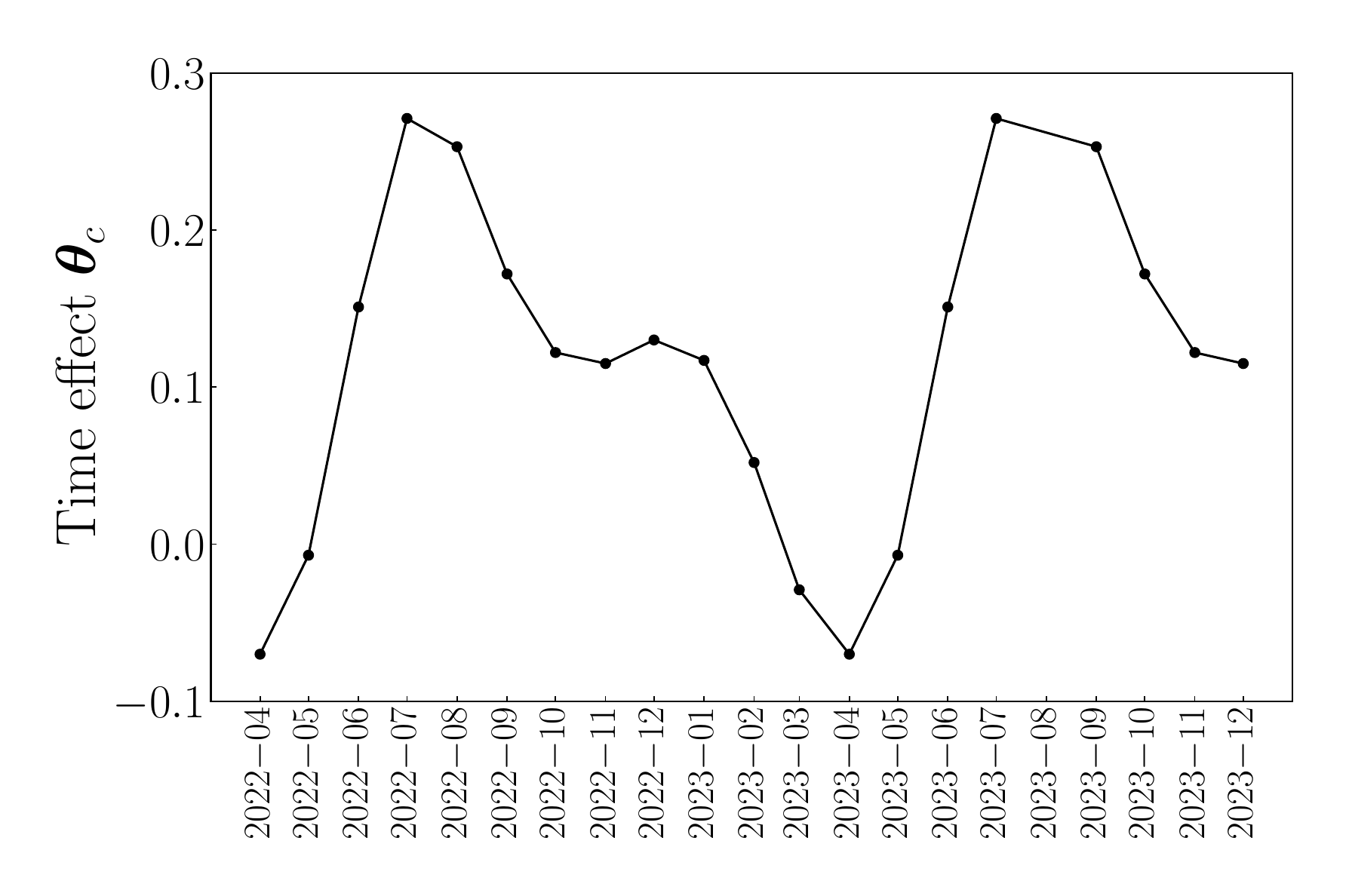}
\caption{Estimated time effect. }
\label{fig:time-effect}
\end{figure}

Figure \ref{fig:random-effects} shows box plots of the mean values for merchants and business types, calculated from the posterior distributions of the estimated merchant and business type effects. 
As shown in Figure \ref{fig:random-effects}, since the dataset comprises different experiments, there is between-study heterogeneity, and the effects vary depending on the merchant and business type. 
This result suggests that modeling random effects using a hierarchical Bayesian approach is appropriate.
The coefficients for the business type effect are $-0.108$ for restaurants and $0.284$ for retail, with the value being higher for retail. 
These results indicate that the promotions encourage consumer stockpiling, as reported in previous studies~\citep[e.g.][]{ailawadi2007decomposition, mela1998long}, and this effect may have appeared more prominently in the retail, where stockpiling is relatively easier.
Figure \ref{fig:time-effect} also shows the estimated time effects arranged chronologically. 
As shown in Figure \ref{fig:time-effect}, there is temporal heterogeneity in the dataset: the effect of promotions varies depending on the implementation period, with the highest effect in July and the lowest in April.
These results suggest that marketers are likely to implement more effective promotions if they carry them out during the summer.

\section{Concluding Remarks}\label{sec:conclusion}
This paper proposes a new Bayesian estimation method called \textit{Shrinkage estimation with Time-trend and Random-Effects for meta-Analysis of Means and variance}~(STREAM). 
This method performs shrinkage estimation of experiments' means and variances using a hierarchical Bayesian approach while accounting for time effects with a Gaussian process. 
The proposed method links related experiments by estimating the shrinkage of means and variances via the hierarchical Bayesian approach, enabling stable estimation of each experiment's parameters through ``borrowing strength.''
Additionally, STREAM flexibly addresses complex time trends in the datasets using a Gaussian process.
Through simulation experiments, we confirmed that STREAM consistently outperforms other baseline methods in both point and interval estimations when between-study heterogeneity is high, regardless of temporal heterogeneity.
To demonstrate the effectiveness of our model, we applied STREAM to a real-world promotion dataset. 
STREAM performed the best in point and interval estimation compared to other baseline methods and showed good convergence in MCMC.

In this study, we assumed a normal distribution for the random effects to represent between-study heterogeneity in the hierarchical Bayesian model. 
However, assuming other distributions may enable more accurate estimation. 
When between-study heterogeneity is high, introducing subgroups as dummy variables relative to the number of experiments in the dataset increases dimensionality, making dimensionality reduction important. 
For instance, dimensionality reduction can be achieved by employing a Laplace distribution~\citep{park2008bayesian} or Horseshoe prior~\citep{carvalho2009handling, carvalho2010horseshoe, qi2022bayesian} as prior distributions for random effects.

In this study, we employed the Gaussian process with a periodic kernel to represent temporal heterogeneity; however, incorporating other kernel functions could potentially achieve more accurate estimation. 
\citet{williams2006gaussian} demonstrated that the sum and product of multiple kernel functions result in kernel functions that maintain positive semi-definiteness. 
It is possible to flexibly address more complex time trends using kernel functions composed of combining multiple kernel functions. 
Exploring methods to achieve higher precision in estimation through comparisons of prior distributions for random effects and comparisons of kernel functions in the Gaussian process within hierarchical Bayesian models remains a future challenge.

\section*{Data availability}
The data that has been used is confidential.

\section*{Competing interests}
All authors report that NTT DOCOMO, INC provided financial support. Kohsuke Kubota and Keiichi Ochiai report a relationship with NTT DOCOMO, INC. that includes: employment. Shonosuke Sugasawa and Takahiro Hoshino report a relationship with NTT DOCOMO, INC. that includes: consulting or advisory.

\bibliographystyle{chicago}
\bibliography{ref}

\begin{thebibliography}{}

\bibitem[\protect\citeauthoryear{Ailawadi, Gedenk, Lutzky, and Neslin}{Ailawadi et~al.}{2007}]{ailawadi2007decomposition}
Ailawadi, K.~L., K.~Gedenk, C.~Lutzky, and S.~A. Neslin (2007).
\newblock Decomposition of the sales impact of promotion-induced stockpiling.
\newblock {\em Journal of Marketing Research\/}~{\em 44\/}(3), 450--467.

\bibitem[\protect\citeauthoryear{Babapulle, Joseph, Bélisle, Brophy, and Eisenberg}{Babapulle et~al.}{2004}]{babapulle2004hierarchical}
Babapulle, M.~N., L.~Joseph, P.~Bélisle, J.~M. Brophy, and M.~J. Eisenberg (2004).
\newblock A hierarchical bayesian meta-analysis of randomised clinical trials of drug-eluting stents.
\newblock {\em The Lancet\/}~{\em 364\/}(9434), 583--591.

\bibitem[\protect\citeauthoryear{Blut, Teller, and Floh}{Blut et~al.}{2018}]{blut2018testing}
Blut, M., C.~Teller, and A.~Floh (2018).
\newblock Testing retail marketing-mix effects on patronage: A meta-analysis.
\newblock {\em Journal of Retailing\/}~{\em 94\/}(2), 113--135.

\bibitem[\protect\citeauthoryear{Borenstein, Hedges, Higgins, and Rothstein}{Borenstein et~al.}{2010}]{borenstein2010basic}
Borenstein, M., L.~V. Hedges, J.~P. Higgins, and H.~R. Rothstein (2010).
\newblock A basic introduction to fixed-effect and random-effects models for meta-analysis.
\newblock {\em Research Synthesis Methods\/}~{\em 1\/}(2), 97--111.

\bibitem[\protect\citeauthoryear{Bowater and Escarela}{Bowater and Escarela}{2013}]{bowater2013heterogeneity}
Bowater, R.~J. and G.~Escarela (2013).
\newblock Heterogeneity and study size in random-effects meta-analysis.
\newblock {\em Journal of Applied Statistics\/}~{\em 40\/}(1), 2--16.

\bibitem[\protect\citeauthoryear{Brahim-Belhouari and Bermak}{Brahim-Belhouari and Bermak}{2004}]{brahim2004gaussian}
Brahim-Belhouari, S. and A.~Bermak (2004).
\newblock Gaussian process for nonstationary time series prediction.
\newblock {\em Computational Statistics \& Data Analysis\/}~{\em 47\/}(4), 705--712.

\bibitem[\protect\citeauthoryear{Browne and Jones}{Browne and Jones}{2017}]{browne2017works}
Browne, W. and M.~S. Jones (2017).
\newblock What works in e-commerce-a meta-analysis of 6700 online experiments.
\newblock {\em Qubit Digital Ltd\/}~{\em 21}.

\bibitem[\protect\citeauthoryear{Carpenter, Gelman, Hoffman, Lee, Goodrich, Betancourt, Brubaker, Guo, Li, and Riddell}{Carpenter et~al.}{2017}]{carpenter2017stan}
Carpenter, B., A.~Gelman, M.~D. Hoffman, D.~Lee, B.~Goodrich, M.~Betancourt, M.~A. Brubaker, J.~Guo, P.~Li, and A.~Riddell (2017).
\newblock Stan: A probabilistic programming language.
\newblock {\em Journal of statistical software\/}~{\em 76}.

\bibitem[\protect\citeauthoryear{Carvalho, Polson, and Scott}{Carvalho et~al.}{2009}]{carvalho2009handling}
Carvalho, C.~M., N.~G. Polson, and J.~G. Scott (2009, 16--18 Apr).
\newblock Handling sparsity via the horseshoe.
\newblock In D.~van Dyk and M.~Welling (Eds.), {\em Proceedings of the Twelfth International Conference on Artificial Intelligence and Statistics}, Volume~5 of {\em Proceedings of Machine Learning Research}, Hilton Clearwater Beach Resort, Clearwater Beach, Florida USA, pp.\  73--80. PMLR.

\bibitem[\protect\citeauthoryear{Carvalho, Polson, and Scott}{Carvalho et~al.}{2010}]{carvalho2010horseshoe}
Carvalho, C.~M., N.~G. Polson, and J.~G. Scott (2010, 04).
\newblock {The horseshoe estimator for sparse signals}.
\newblock {\em Biometrika\/}~{\em 97\/}(2), 465--480.

\bibitem[\protect\citeauthoryear{Chowdhry, Dworkin, and McDermott}{Chowdhry et~al.}{2016}]{chowdhry2016meta}
Chowdhry, A.~K., R.~H. Dworkin, and M.~P. McDermott (2016).
\newblock Meta-analysis with missing study-level sample variance data.
\newblock {\em Statistics in Medicine\/}~{\em 35\/}(17), 3021--3032.

\bibitem[\protect\citeauthoryear{Dakin, Welton, Ades, Collins, Orme, and Kelly}{Dakin et~al.}{2011}]{dakin2011mixed}
Dakin, H.~A., N.~J. Welton, A.~Ades, S.~Collins, M.~Orme, and S.~Kelly (2011).
\newblock Mixed treatment comparison of repeated measurements of a continuous endpoint: an example using topical treatments for primary open-angle glaucoma and ocular hypertension.
\newblock {\em Statistics in Medicine\/}~{\em 30\/}(20), 2511--2535.

\bibitem[\protect\citeauthoryear{Duane, Kennedy, Pendleton, and Roweth}{Duane et~al.}{1987}]{duane1987hybrid}
Duane, S., A.~Kennedy, B.~J. Pendleton, and D.~Roweth (1987).
\newblock Hybrid monte carlo.
\newblock {\em Physics Letters B\/}~{\em 195\/}(2), 216--222.

\bibitem[\protect\citeauthoryear{Gelman}{Gelman}{2006}]{gelman2006prior}
Gelman, A. (2006).
\newblock Prior distributions for variance parameters in hierarchical models (comment on an article by browne and draper).
\newblock {\em Bayesian Analysis\/}~{\em 1}, 515--533.

\bibitem[\protect\citeauthoryear{Gelman, Carlin, Stern, Dunson, Vehtari, and Rubin}{Gelman et~al.}{2013}]{gelman2013bayesian}
Gelman, A., J.~B. Carlin, H.~S. Stern, D.~B. Dunson, A.~Vehtari, and D.~B. Rubin (2013).
\newblock {\em Bayesian Data Analysis\/} (3rd ed.).
\newblock Chapman and Hall/CRC.

\bibitem[\protect\citeauthoryear{Gelman and Rubin}{Gelman and Rubin}{1992}]{gelman1992inference}
Gelman, A. and D.~B. Rubin (1992).
\newblock {Inference from Iterative Simulation Using Multiple Sequences}.
\newblock {\em Statistical Science\/}~{\em 7\/}(4), 457 -- 472.

\bibitem[\protect\citeauthoryear{Gene~Hwang, Qiu, and Zhao}{Gene~Hwang et~al.}{2009}]{gene2009empirical}
Gene~Hwang, J., J.~Qiu, and Z.~Zhao (2009).
\newblock Empirical bayes confidence intervals shrinking both means and variances.
\newblock {\em Journal of the Royal Statistical Society Series B: Statistical Methodology\/}~{\em 71\/}(1), 265--285.

\bibitem[\protect\citeauthoryear{Gneiting and Raftery}{Gneiting and Raftery}{2007}]{gneiting2007strictly}
Gneiting, T. and A.~E. Raftery (2007).
\newblock Strictly proper scoring rules, prediction, and estimation.
\newblock {\em Journal of the American statistical Association\/}~{\em 102\/}(477), 359--378.

\bibitem[\protect\citeauthoryear{Higgins, Thompson, and Spiegelhalter}{Higgins et~al.}{2009}]{higgins2009re}
Higgins, J.~P., S.~G. Thompson, and D.~J. Spiegelhalter (2009).
\newblock A re-evaluation of random-effects meta-analysis.
\newblock {\em Journal of the Royal Statistical Society Series A: Statistics in Society\/}~{\em 172\/}(1), 137--159.

\bibitem[\protect\citeauthoryear{Kleinbaum, Kupper, and Morgenstern}{Kleinbaum et~al.}{1991}]{kleinbaum1991epidemiologic}
Kleinbaum, D.~G., L.~L. Kupper, and H.~Morgenstern (1991).
\newblock {\em Epidemiologic research: principles and quantitative methods}.
\newblock John Wiley \& Sons.

\bibitem[\protect\citeauthoryear{LeBlanc and Banks}{LeBlanc and Banks}{2022}]{leblanc2022time}
LeBlanc, P.~M. and D.~Banks (2022).
\newblock Time-varying bayesian network meta-analysis.
\newblock {\em arXiv preprint arXiv:2211.08312\/}.

\bibitem[\protect\citeauthoryear{Lindley and Smith}{Lindley and Smith}{1972}]{lindley1972bayes}
Lindley, D.~V. and A.~F.~M. Smith (1972).
\newblock Bayes estimates for the linear model.
\newblock {\em Journal of the Royal Statistical Society. Series B (Methodological)\/}~{\em 34\/}(1), 1--41.

\bibitem[\protect\citeauthoryear{Liu-Thompkins, Khoshghadam, {Attar Shoushtari}, and Zal}{Liu-Thompkins et~al.}{2022}]{liuthompkins2022what}
Liu-Thompkins, Y., L.~Khoshghadam, A.~{Attar Shoushtari}, and S.~Zal (2022).
\newblock What drives retailer loyalty? a meta-analysis of the role of cognitive, affective, and social factors across five decades.
\newblock {\em Journal of Retailing\/}~{\em 98\/}(1), 92--110.

\bibitem[\protect\citeauthoryear{Mawdsley, Bennetts, Dias, Boucher, and Welton}{Mawdsley et~al.}{2016}]{mawdsley2016model}
Mawdsley, D., M.~Bennetts, S.~Dias, M.~Boucher, and N.~J. Welton (2016).
\newblock Model-based network meta-analysis: a framework for evidence synthesis of clinical trial data.
\newblock {\em CPT: pharmacometrics \& systems pharmacology\/}~{\em 5\/}(8), 393--401.

\bibitem[\protect\citeauthoryear{Mela, Jedidi, and Bowman}{Mela et~al.}{1998}]{mela1998long}
Mela, C.~F., K.~Jedidi, and D.~Bowman (1998).
\newblock The long-term impact of promotions on consumer stockpiling behavior.
\newblock {\em Journal of Marketing Research\/}~{\em 35\/}(2), 250--262.

\bibitem[\protect\citeauthoryear{Miller and Hosanagar}{Miller and Hosanagar}{2020}]{miller2020empirical}
Miller, A.~P. and K.~Hosanagar (2020).
\newblock An empirical meta-analysis of e-commerce {A}/{B} testing strategies.
\newblock Technical report, Working paper.

\bibitem[\protect\citeauthoryear{Oehlert}{Oehlert}{1992}]{oehlert1992note}
Oehlert, G.~W. (1992).
\newblock A note on the delta method.
\newblock {\em The American Statistician\/}~{\em 46\/}(1), 27--29.

\bibitem[\protect\citeauthoryear{Park and Casella}{Park and Casella}{2008}]{park2008bayesian}
Park, T. and G.~Casella (2008).
\newblock The bayesian lasso.
\newblock {\em Journal of the American Statistical Association\/}~{\em 103\/}(482), 681--686.

\bibitem[\protect\citeauthoryear{Pastor and Lazowski}{Pastor and Lazowski}{2018}]{pastor2018on}
Pastor, D.~A. and R.~A. Lazowski (2018).
\newblock On the multilevel nature of meta-analysis: A tutorial, comparison of software programs, and discussion of analytic choices.
\newblock {\em Multivariate Behavioral Research\/}~{\em 53\/}(1), 74--89.

\bibitem[\protect\citeauthoryear{Pedder, Dias, Bennetts, Boucher, and Welton}{Pedder et~al.}{2019}]{pedder2019modelling}
Pedder, H., S.~Dias, M.~Bennetts, M.~Boucher, and N.~J. Welton (2019).
\newblock Modelling time-course relationships with multiple treatments: Model-based network meta-analysis for continuous summary outcomes.
\newblock {\em Research Synthesis Methods\/}~{\em 10\/}(2), 267--286.

\bibitem[\protect\citeauthoryear{Qi, Zhou, Wang, and Peterson}{Qi et~al.}{2022}]{qi2022bayesian}
Qi, X., S.~Zhou, Y.~Wang, and C.~Peterson (2022).
\newblock Bayesian sparse modeling to identify high-risk subgroups in meta-analysis of safety data.
\newblock {\em Research Synthesis Methods\/}~{\em 13\/}(6), 807--820.

\bibitem[\protect\citeauthoryear{Roberts, Osborne, Ebden, Reece, Gibson, and Aigrain}{Roberts et~al.}{2013}]{roberts2013gaussian}
Roberts, S., M.~Osborne, M.~Ebden, S.~Reece, N.~Gibson, and S.~Aigrain (2013).
\newblock Gaussian processes for time-series modelling.
\newblock {\em Philosophical Transactions of the Royal Society A: Mathematical, Physical and Engineering Sciences\/}~{\em 371\/}(1984), 20110550.

\bibitem[\protect\citeauthoryear{Salanti, Marinho, and Higgins}{Salanti et~al.}{2009}]{salanti2009standard}
Salanti, G., V.~Marinho, and J.~P. Higgins (2009).
\newblock A case study of multiple-treatments meta-analysis demonstrates that covariates should be considered.
\newblock {\em Journal of Clinical Epidemiology\/}~{\em 62\/}(8), 857--864.

\bibitem[\protect\citeauthoryear{Sohn}{Sohn}{1999}]{sohn1999meta}
Sohn, S.~Y. (1999).
\newblock Meta analysis of classification algorithms for pattern recognition.
\newblock {\em IEEE Trans. Pattern Anal. Mach. Intell.\/}~{\em 21\/}(11), 1137–1144.

\bibitem[\protect\citeauthoryear{Stanley and Doucouliagos}{Stanley and Doucouliagos}{2012}]{stanley2012meta}
Stanley, T.~D. and H.~Doucouliagos (2012).
\newblock {\em Meta-regression analysis in economics and business}.
\newblock routledge.

\bibitem[\protect\citeauthoryear{Stourm and Bradlow}{Stourm and Bradlow}{2023}]{stourm2023cross}
Stourm, V. and E.~T. Bradlow (2023).
\newblock Cross-reward effects in a coalition loyalty program: The impact of a point currency devaluation.
\newblock {\em International Journal of Research in Marketing\/}~{\em 40\/}(2), 276--293.

\bibitem[\protect\citeauthoryear{Sugasawa and Kubokawa}{Sugasawa and Kubokawa}{2020}]{sugasawa2020small}
Sugasawa, S. and T.~Kubokawa (2020).
\newblock Small area estimation with mixed models: a review.
\newblock {\em Japanese Journal of Statistics and Data Science\/}~{\em 3}, 693--720.

\bibitem[\protect\citeauthoryear{Sugasawa, Tamae, and Kubokawa}{Sugasawa et~al.}{2017}]{sugasawa2017bayesian}
Sugasawa, S., H.~Tamae, and T.~Kubokawa (2017).
\newblock Bayesian estimators for small area models shrinking both means and variances.
\newblock {\em Scandinavian Journal of Statistics\/}~{\em 44\/}(1), 150--167.

\bibitem[\protect\citeauthoryear{Thompson and Becker}{Thompson and Becker}{2020}]{thompson2020group}
Thompson, C.~G. and B.~J. Becker (2020).
\newblock A group-specific prior distribution for effect-size heterogeneity in meta-analysis.
\newblock {\em Behavior research methods\/}~{\em 52}, 2020--2030.

\bibitem[\protect\citeauthoryear{van Enst, Naaktgeboren, Ochodo, de~Groot, Leeflang, Reitsma, Scholten, Moons, Zwinderman, Bossuyt, and Hooft}{van Enst et~al.}{2015}]{vanEnst2015small}
van Enst, W.~A., C.~A. Naaktgeboren, E.~A. Ochodo, J.~A.~H. de~Groot, M.~M.~G. Leeflang, J.~B. Reitsma, R.~J. P.~M. Scholten, K.~G.~M. Moons, A.~H. Zwinderman, P.~M.~M. Bossuyt, and L.~Hooft (2015).
\newblock Small-study effects and time trends in diagnostic test accuracy meta-analyses: a meta-epidemiological study.
\newblock {\em Systematic Reviews\/}~{\em 4}, 66.

\bibitem[\protect\citeauthoryear{Wang, Lin, Hodges, MacLehose, and Chu}{Wang et~al.}{2021}]{wang2021variance}
Wang, Z., L.~Lin, J.~S. Hodges, R.~MacLehose, and H.~Chu (2021).
\newblock A variance shrinkage method improves arm-based bayesian network meta-analysis.
\newblock {\em Statistical methods in medical research\/}~{\em 30\/}(1), 151--165.

\bibitem[\protect\citeauthoryear{Williams and Rasmussen}{Williams and Rasmussen}{2006}]{williams2006gaussian}
Williams, C.~K. and C.~E. Rasmussen (2006).
\newblock {\em Gaussian processes for machine learning}.
\newblock MIT press Cambridge, MA.

\bibitem[\protect\citeauthoryear{Williams, Rodriguez, and Bürkner}{Williams et~al.}{2021}]{williams2021putting}
Williams, D.~R., J.~E. Rodriguez, and P.~C. Bürkner (2021).
\newblock Putting variation into variance: Modeling between-study heterogeneity in meta-analysis.
\newblock PsyArXiv.

\bibitem[\protect\citeauthoryear{Zerbini, Bijmolt, Maestripieri, and Luceri}{Zerbini et~al.}{2022}]{zerbini2022drivers}
Zerbini, C., T.~H. Bijmolt, S.~Maestripieri, and B.~Luceri (2022).
\newblock Drivers of consumer adoption of e-commerce: A meta-analysis.
\newblock {\em International Journal of Research in Marketing\/}~{\em 39\/}(4), 1186--1208.

\end{thebibliography}

\end{document}